\documentclass[12pt]{article}

\def\theorem #1. #2\par{\medbreak
  \noindent{\tt {\bf Theorem  #1.}\enspace}{\sl#2\par}%
  \ifdim\lastskip<\medskipamount \removelastskip\penalty55\medskip\fi}

\def\{{\lbrace}
\def\}{\rbrace}

\def\cl{{\cal C}\!\ell}
\def\ra{\rangle}

\def\ss{\stackrel}

\def\n{{\scriptsize\textsc n}}
\def\k{{\scriptsize\textsc k}}

\def\ww{\wedge\ldots\wedge}

\def\I{{\cal I}}
\def\R{{\cal R}}
\def\M{{\cal M}}
\def\C{{\cal C}}

\def\diag{{\rm diag}}

\def\be{\begin{equation}}
\def\ee{\end{equation}}

\begin{document}
\setcounter{page}{1}
\baselineskip=16pt

\author{N.G.Marchuk}

\title{A coordinateless form of the Dirac equation}

\maketitle

Steklov Mathematical Institute, Gubkina st.8, Moscow 119991, Russia

nmarchuk@mi.ras.ru,   www.orc.ru/\~{}nmarchuk

\begin{abstract}
We present a so called Dirac-type tensor equation (DTTE). This equation
is written in coordinateless form with the aid of differential operators
$d$ and $\delta$. A wave function of DTTE  belongs to a minimal left
ideal of the algebra of exterior forms with respect to the Clifford
product. We show that a coordinate form of DTTE is identical to the
Dirac equation in a fixed coordinate system.
\end{abstract}


Let $\R^{1,3}$  be the Minkowski space with coordinates $x^\mu$ ($\mu=0,1,2,3$)
and $\eta$ be the Minkowski matrix
$\eta=\|\eta_{\mu\nu}\|=\|\eta^{\mu\nu}\|=\diag(1,-1,-1,-1)$.
The Dirac equation for an electron has the form
\begin{equation}
\gamma^\mu(\partial_\mu\psi+ia_\mu\psi)+im\psi=0,
\label{Dirac:eq}
\end{equation}
where $\gamma^\mu$ are complex valued
matrices such that
$$
\gamma^\mu\gamma^\nu+\gamma^\nu\gamma^\mu=2\eta^{\mu\nu}{\bf 1},
\quad \psi=\psi(x)=(\psi^1,\ldots,\psi^4)^{\rm T}
$$
and ${\bf 1}$ is the identity matrix of fourth order. Summation convention
over repeating indices is assumed and
$\partial_\mu=\partial/\partial
x^\mu$.

In 1962 E.K\"ahler
\cite{Kahler} suggests the following equation for the electron
(the Dirac-K\"ahler equation)
$$
(d-\delta)\Phi+iA\Phi+im\Phi=0,\quad \Phi\in\Lambda,
\quad A\in\Lambda_1^\R.
$$
The equation is written in invariant (coordinateless) form.
Several types of algebra-geometric objects are used in this equation.

\noindent {\bf Exterior forms.} Suppose that in the Minkowski space
$\R^{1,3}$ with coordinates
$x^\mu$ we have basis coordinates vectors
$e_\mu$ and basis covectors
$e^\mu=\eta^{\mu\nu}e_\nu$.
Let $\Lambda_k$ be the set of exterior $k$-forms
$\frac{1}{k!}\phi_{\mu_1\ldots\mu_k}e^{\mu_1}\ww e^{\mu_k}
\in\Lambda_k$, where $\phi_{\mu_1\ldots\mu_k}=\phi_{[\mu_1\ldots\mu_k]}$
are complex valued antisymmetric covariant rang-$k$ tensors.
We take $\Lambda=\Lambda_0\oplus\ldots\oplus\Lambda_4$.
Complex dimensions of the spaces
$\Lambda_k$ and
$\Lambda$ are equal to $1,4,6,4,1$ and 16 respectively.

\noindent{\bf Hodge star operation}
$\star\,:\,\Lambda_k\to\Lambda_{4-k}$ identifies  $\Lambda_k$ with
$\Lambda_{4-k}$
$$
\star\ss{k}{\Phi}=
\frac{1}{k!(4-k)!}\varepsilon_{\mu_1\ldots\mu_4}\phi^{\mu_1\ldots\mu_k}
e^{\mu_{k+1}}\ww e^{\mu_4},\quad
\star\star\ss{k}{\Phi}=(-1)^{k+1}\ss{k}{\Phi},
$$
where $\varepsilon_{\mu_1\ldots\mu_4}$ is the completely antisymmetric tensor
and $\varepsilon_{0123}=1$.

\noindent {\bf The differential} $d:\Lambda_k\to\Lambda_{k+1}$
$$
d\Phi=e^\mu\wedge\partial_\mu\Phi, \quad d^2=0.
$$

\noindent {\bf Codifferential} $\delta:\Lambda_k\to\Lambda_{k-1}$
$$
\delta=\star d\star,\quad \delta^2=0.
$$

\noindent {\bf The Clifford product} $UV\in\Lambda$ of exterior forms
$U,V\in\Lambda$ is defined by the rules
\begin{description}
\item[(i)] $\Lambda$ is an associative algebra with identity element $1$
with respect to the Clifford product.
\item[(ii)] $e^\mu e^\nu=e^\mu\wedge e^\nu+\eta^{\mu\nu}$.
\item[(iii)] $e^{\mu_1}\ldots e^{\mu_k}=e^{\mu_1}\ww e^{\mu_k}\quad
\hbox{for}\quad \mu_1<\cdots<\mu_k$.
\end{description}

It follows from this rules that
\begin{description}
\item[1)]
$
e^\mu e^\nu+e^\nu e^\mu=2\eta^{\mu\nu}.
$
\item[2)] $
A^2=a_\mu a^\mu\in\Lambda_0  \quad\hbox{for}\quad A=a_\mu e^\mu\in\Lambda_1,
$
\item[3)] $
A\Phi=A\wedge\Phi-\star(A\wedge\star\Phi)
\quad\hbox{for}\quad A\in\Lambda_1,\Phi\in\Lambda.
$
\end{description}

The identity 1) is the main identity in the Clifford algebra
$\cl(1,3)$.

The operations of differential, codifferential, and Clifford product
take tensors (exterior forms) to tensors. Hence the Dirac-K\"ahler
equation is a tensor equation.

Let us take a set of four linear independent orthonormal covectors
$h_\mu{}^a$. This set is called a {\em tetrad}.
$$
h_\mu{}^a h^{\mu b}=\eta^{ab},\quad \partial_\mu h_\nu{}^a=0.
$$
Also we take tetrad 1-forms
$h^a=h_\mu{}^a e^\mu$.
We use Greed indices as tensorial and Latin indices as nontensorial
(tetrad) indices. We raise and lower  Greek and Latin indices with the
aid of the Minkowski matrix
$h_{\nu a}=\eta_{ab}h_\nu{}^b$, $h^{\nu a}=\eta^{\mu\nu}h_\mu{}^a$.

Consider the operation of Clifford conjugation
$*\,:\,\Lambda_k\to\Lambda_k$. For
$U\in\Lambda_k$
$$
U^*=(-1)^{\frac{k(k-1)}{2}}\bar{U},
$$
where $\bar{U}$ is the exterior form with complex conjugated components.

Denote $E=h^0$. Now we can define the operation of Hermitian conjugation
of exterior forms
$\dagger\,:\,\Lambda_k\to\Lambda_k$
$$
U^\dagger=E U^* E \quad\hbox{for}\quad U\in\Lambda.
$$
Evidently,
$$
(UV)^\dagger=V^\dagger U^\dagger,\quad U^{\dagger\dagger}=U,
\quad i^\dagger=-i.
$$
An exterior form
$U\in\Lambda$ is called {\em Hermitian} if
$U^\dagger=U$, and anti-Hermitian if $U^\dagger=-U$.

An element $t$ of an algebra is called {\em an idempotent} (or
projector) if
$t^2=t$. Any idempotent generates the left ideal. A primitive idempotent
generates the minimal left ideal.

Let
$t\in\Lambda$ be an idempotent of
$\Lambda$ with respect to Clifford product such that
$$
t^2=t,\quad \partial_\mu t=0, \quad t^\dagger=t
$$
and
$$
\I(t)=\{Ut\,:\,U\in\Lambda\}
$$
be the minimal left ideal of $\Lambda$.

Let us introduce {\em a Dirac-type tensor equation} (DTTE) in
coordinateless form
\begin{equation}
(d-\delta)\Psi+iA\Psi+im\Psi=0,\quad \Psi\in\I(t),\quad
A\in\Lambda_1^\R.
\label{Dtt0}
\end{equation}
The exterior form $\Psi$ in DTTE belongs to the minimal left ideal
$\I(t)\subset\Lambda$. And this is the only difference between the
Dirac-K\"ahler equation and DTTE.
In the sequel we connect DTTE with the Dirac equation
(\ref{Dirac:eq}).

Recall that M.~Riesz
in 1947 was the first who consider Dirac spinors as
elements of a minimal left ideal of the Clifford algebra (a partial case
of pure spinors was considered by E.~Cartan in 1938). M.~Riesz deals
with the spinor representation of the Dirac equation.

Now we want to consider the coordinate form of DTTE instead of the
coordinateless form. For this we take the idempotent
$$
t=\frac{1}{4}(1+h^0)(1+ih^1 h^2)\in\Lambda
$$
such that $\quad t^2=t$, $\partial_\mu t=0$, $t^\dagger=t$.
It is easy to check that the following 1-forms:
\begin{equation}
t_1=t,\quad t_2=-h^1 h^3 t,\quad t_3=h^0 h^3 t,\quad t_4=h^0 h^1 t
\label{tttt}
\end{equation}
form a basis of the minimal left ideal
$\I(t)$. With the aid of this basis we define the map
$\gamma:\Lambda\to \M(4,\C)$, where $\M(4,\C)$ is the algebra
of complex valued matrices of fourth order. Namely, for
$U\in\Lambda$
$$
Ut_\n=\gamma(U)^\k_\n t_\k,
$$
where $\gamma(U)^\k_\n$ are the elements of the matrix $\gamma(U)$
(an upper index enumerates rows and a lower index enumerates columns)
Note that we take the basis
(\ref{tttt}) such that
$$
\gamma(U^\dagger)=(\gamma(U))^\dagger,
$$
where $U^\dagger$ is the Hermitian conjugated form $U^\dagger=EU^*E$,
and
$(\gamma(U))^\dagger$ is the Hermitian conjugated matrix
(transposed matrix with complex conjugated elements).

Also we use the map that takes elements of the left ideal
$\Omega\in\I(t)$ to ket vector
$|\Omega\ra\in \C^4$.
If $\Omega=\omega^\k t_\k\in\I(t)$, then
$$
|\Omega\ra=(\omega^1,\ldots, \omega^4)^{\rm T}.
$$
For $U,V\in\Lambda,\Omega\in\I(t)$ we have
$$
|U\Omega\ra=\gamma(U)|\Omega\ra, \quad
\gamma(UV)=\gamma(U)\gamma(V),
$$
i.e, $\gamma$ is a matrix representation of elements of
$\Lambda$.
It can be shown that
$$
d-\delta=e^\mu\partial_\mu.
$$
Let us rewrite DTTE in the form
\begin{equation}
\Omega\equiv e^\mu(\partial_\mu\Psi+ia_\mu\Psi)+im\Psi=0
\label{'""'}
\end{equation}
and get the ket vector
\begin{equation}
|\Omega\ra=\gamma^\mu(\partial_\mu\psi+ia_\mu\psi)+im\psi=0,
\label{Omega1}
\end{equation}
where $\psi=|\Psi\ra$,
$\gamma^\mu=\gamma(e^\mu)$.
From the identity
$e^\mu e^\nu+e^\nu e^\mu=2\eta^{\mu\nu}$ we have
$\gamma^\mu\gamma^\nu+\gamma^\nu\gamma^\mu=2\eta^{\mu\nu}{\bf 1}$.

Hence we get the theorem.

\theorem. The coordinate form (\ref{Omega1}) of DTTE coincides with
the Dirac equation
(\ref{Dirac:eq}).\par


\end{document}